\documentclass{article}

\usepackage{arxiv}

\usepackage[utf8]{inputenc} 
\usepackage[T1]{fontenc}    
\usepackage{hyperref}       
\usepackage{url}            
\usepackage{booktabs}       
\usepackage{amsfonts}       
\usepackage{nicefrac}       
\usepackage{microtype}      
\usepackage{graphicx}
\usepackage{doi}

\usepackage{adjustbox}
\usepackage{braket}
\usepackage{caption}
\usepackage{mathtools}
\usepackage{multirow}
\usepackage{stackengine}
\usepackage{tabularx}
\usepackage{verbatim}
\usepackage[table]{xcolor}

\DeclareMathOperator{\Tr}{Tr}
\newcolumntype{Y}{>{\centering\arraybackslash}X}

\hypersetup{
    unicode=true,
	colorlinks,
	citecolor=black,
	filecolor=black,
	linkcolor=black,
	urlcolor=black,
}

\title{Subdiffusive semantic evolution\texorpdfstring{\\}{ }in Indo-European languages}

\date{\today}	

\author{ \href{https://orcid.org/0000-0002-1085-1939}{\includegraphics[scale=0.06]{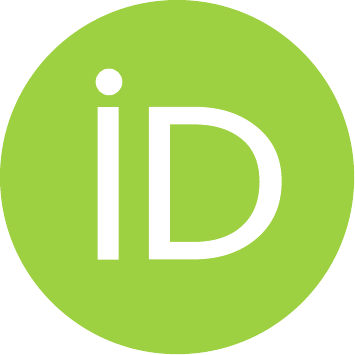}\hspace{1mm}Bogdán Asztalos}
        \\
	Dept.\ of Biological Physics, E\"{o}tv\"{o}s University, H-1117 Budapest, P{\'a}zm{\'a}ny P. stny. 1/A, Hungary \\
	\AND
	\href{https://orcid.org/0000-0002-3406-4200}{\includegraphics[scale=0.06]{orcid.pdf}\hspace{1mm}Gergely Palla}\\
	Dept.\ of Biological Physics, E\"{o}tv\"{o}s University, H-1117 Budapest, P{\'a}zm{\'a}ny P. stny. 1/A, Hungary, \\
	Health Services Management Training Centre, Semmelweis University, H-1125 Budapest, K{\'u}tv{\"o}lgyi {\'u}t 2, Hungary
	\AND
	\href{https://orcid.org/0000-0002-5722-1598}{\includegraphics[scale=0.06]{orcid.pdf}\hspace{1mm}Dániel Czégel}\\
	Beyond Center for Fundamental Concepts in Science, Arizona State University, Tempe, AZ 85287, USA \\
	Institute of Evolution, Centre for Ecological Research, H-1121 Budapest, Hungary \\
	\texttt{dczegel@asu.edu} \\
}

\renewcommand{\shorttitle}{Subdiffusive semantic evolution in Indo-European languages}

\hypersetup{
pdftitle={Subdiffusive semantic evolution in Indo-European languages},
pdfsubject={},
pdfauthor={Bogdán Asztalos, Gergely Palla, Dániel Czégel},
pdfkeywords={semantic evolution, distributional semantics, word embedding, anomalous diffusion},
}

\begin{document}
\maketitle

\begin{abstract} 

How do words change their meaning? Although semantic evolution is driven by a variety of distinct factors, including linguistic, societal, and technological ones, we find that there is one law that holds universally across five major Indo-European languages: that semantic evolution is strongly subdiffusive. Using an automated pipeline of diachronic distributional semantic embedding that controls for underlying symmetries, we show that words follow stochastic trajectories in meaning space with an anomalous diffusion exponent $\alpha= 0.45\pm 0.05$ across languages, in contrast with diffusing particles that follow $\alpha=1$. Randomization methods indicate that preserving temporal correlations in semantic change \emph{directions} is necessary to recover strongly subdiffusive behavior; however, correlations in change \emph{sizes} play an important role too. We furthermore show that strong subdiffusion is a robust phenomenon under a wide variety of choices in data analysis and interpretation, such as the choice of fitting an ensemble average of displacements or averaging best-fit exponents of individual word trajectories.

\end{abstract}


\section{Introduction}
Cumulative cultural evolution at enormous scale and speed makes us a strikingly different species than the rest of the living world \cite{mesoudi2021cultural}. The ability of accumulating techniques and solutions one bit at a time, aggregating them across time and space gave us unprecedented power, that we use and abuse to shape the world. But how is this process of human cultural evolution unfold? Are there universal patterns that hold across cultures and eras? 
One human activity that allows us to zoom into the cumulative patters of human thought is language production. We all comprehend and produce, all the time, keeping the cogwheels of language evolution moving. How do these cogwheels move?

Language has a unique property: it is made of atomic elements, morphemes, minimal units that possess “meaning”, whose definition is well agreed upon. Much of the information in any intricate pattern of text is concentrated on the scales of selecting and combining these atoms in a combinatorial manner, just like much of the information that determines the bauplan of a biological organism is concentrated on the level of genes. 
Studying the dynamics of these elementary units or a subset of them (i.e., units that determine content and not grammatical function), at a large-scale, data-driven way might shed light to basic universal mechanisms of cultural evolutionary processes.

Motivated by this line of reasoning, recent efforts focused on uncovering universal (i.e. language, time, genre, etc. independent) dynamical rules that govern \emph{frequency changes} of (variants of) words or ordered combinations of them, called n-grams \cite{ petersen2012statistical}. This effort provided us a handful of remarkable discoveries, including peaks or valleys of individual terms mirroring specific societal-political processes (e.g., censorship, propaganda, ideology shifts, cultural-technological drift, natural or social catastrophes, etc.), aggregate behavior of groups of terms (e.g., decay rate of fame of a cohort of people across historical time) \cite{michel2011quantitative}, competition dynamics among linguistic variants (e.g., \cite{lieberman2007quantifying}) or synonyms, and even a marked difference between the temporal correlational patterns of word frequencies referring to natural or social processes \cite{gao2012culturomics}.

Language provides us a unique opportunity to peer into abstract cultural processes in another way: in that meaning can be anchored by asking human subjects to \emph{report} about it. This makes it possible to automate the process of meaning extraction (to some approximation) by relating algorithmic outcomes with those of large-scale psycholinguistic experiments, as well as other natural language processing tasks \cite{levy2015improving,lenci2022comparative}. In particular, one data-driven technique for estimating semantic similarity of words has been dominant over the past decades: \emph{distributional semantics}. It builds on the so-called distributional hypothesis, paraphrased as “a word is characterized by the company it keeps”\cite{harris1954distributional}: similarity in meaning can be approximated by comparing neighborhoods of words over large corpora \cite{lenci2018distributional}. 
Many flavors of the distributional hypothesis has been formalized with classical methods \cite{landauer1998introduction,papadimitriou2000latent}, yet the advent of large-scale estimation of semantic similarity came with the ”machine learning revolution” \cite{levy2015improving}. From these techniques, a successful and computationally efficient variant is the Word2vec embedding algorithm \cite{mikolov2013distributed, goldberg2014word2vec}. Word2vec estimates semantic similarity based on sampling co-occurrences of words. It implements dimension reduction over the set of pairwise semantic similarities to embed words in a relatively low dimensional space (e.g., tens of thousands of words in a few hundred dimensional Euclidean space) \cite{levy2014neural}.  With a corpus of time-labelled co-occurrences in hand, one can, in principle, track changes of these semantic similarities in an automated way by comparing embeddings corresponding to different times.

Indeed, a first endeavor utilizing this approach has identified two novel statistical patterns governing the \emph{semantic change} of words: the law of conformity, stating that words with lower frequency change their meaning faster, and the law of innovation, finding that more polysemous words also tend to exhibit higher rate of semantic change \cite{hamilton2016diachronic}. Although the authors applied multiple word embedding variants, the validity of these results mirroring social-technological-linguistic effects as opposed to being mathematical artefacts is still debated \cite{dubossarsky2017outta}. It is because although state of the art word embedding methods match reported semantic similarities across experiments, languages, and training corpora, they produce systematic biases over non-semantic features. One particular bias is due to the power-law distribution of word occurrences, known as Zipf's law \cite{zipf1949human, ryland2015zipf}, resulting in embeddings where low-frequency words tend to appear close to the center of the embedding whereas high-frequency words are pushed to the periphery. Such implicit biases point at the necessity of careful comparison of obtained results with those based on various randomized replicas of the dataset, systematically removing statistical dependencies until the phenomena at hand is no longer apparent.


Furthermore, with \emph{diachronic} embeddings, three additional issues arise. First, embedding dimensions are arbitrary. There is no guarantee that dimensions match across subsequent timesteps. Second, Word2vec is a sampling-based method, and therefore, it is non-deterministic. Third, there are underlying symmetries along which embeddings are degenerate: namely, a set of transformations that change embeddings (and the embedding of context words) yet leaves co-occurrences invariant. In order to tackle all four of aforementioned obstacles, in this paper we develop a dynamical alignment method that is i) symmetry-agnostic, ii) averages over many runs to yield a robust estimate of embedding positions and their variances, iii) based on these principles, finds a best alignment over all timesteps, and iv) compares obtained results with those of systematically randomized versions of input data, removing various statistical dependencies in a step-by-step manner. Current diachronic embedding methods mostly focus on point iii) \cite{bamler2017dynamic, dridi2022vec2dynamics, yao2018dynamic}; here we suggest that all points above are needed for a robust estimation of semantic trajectories.

Although instantaneous statistical patterns of ensembles of words (e.g., positions, velocities) are interesting, there is a fundamental aspect in which evolutionary (be it cultural or biological) processes differ from most physical ones: \emph{historical contingency}. We therefore shift focus from an ensemble of words to an \emph{ensemble of word trajectories} point of view, and ask: are there robust statistical regularities, universal across languages, that govern semantic evolution? Building on a theoretically underpinned pipeline of embedding, temporal alignment, and systematic randomization outlined above, we analyize ensemble of word trajectories using methods from non-equilibrium statistical mechanics. We focus on identifying and explaining systematic deviations from stochastic trajectories of standard diffusive particles, pointing at non-trivial long-range spatiotemporal correlations between word trajectories. 


\section{Results}
\label{section_results}

\subsection{Maximally representation-agnostic temporal embedding} \label{subsec_temporal-embedding}

\begin{figure}
    \centering
    \includegraphics[width=\linewidth]{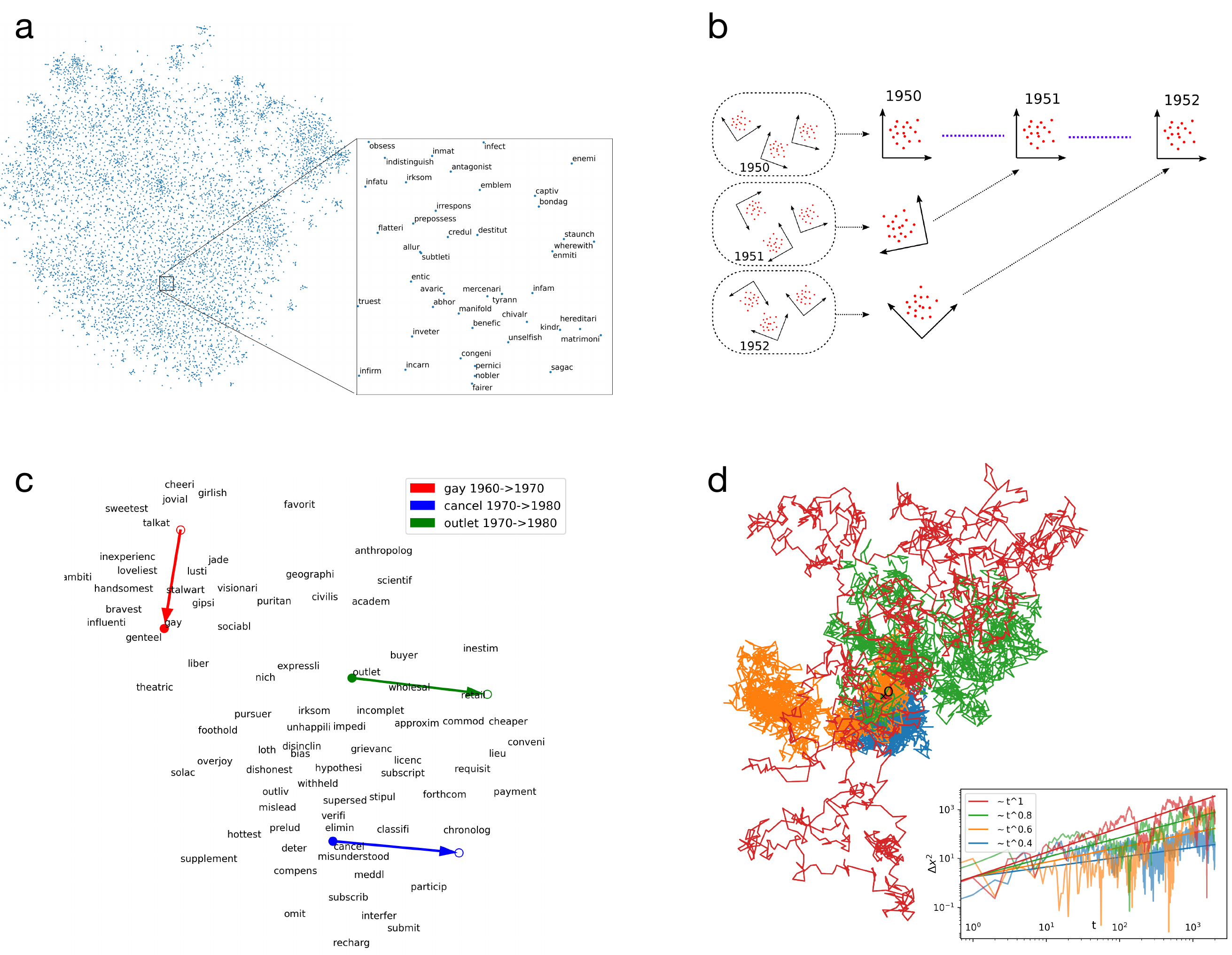}
    \caption{(a) 2D projection of a high-dimensional word embedding, illustrating semantic similarity. 300D embedding of lemmas in the Google ngram English fiction database are generated by Word2vec Skip-gram model and are nonlinearly projected to 2D using t-sne \cite{van2008visualizing}. (b) Alignment of embeddings. First, multiple embeddings of the same year are aligned and averaged to reduce embedding noise, then, these averaged embeddings are aligned across time to achieve maximally smoothened word trajectories. (c) The resulting diachronic embedding makes it possible to visualize semantic change. Three selected words (gay, cancel, outlet) with most change within a decade are shown over a static background. (d) Illustration of subdiffusive trajectories generated by fractional Brownian motion. Inset: mean squared distance scales as $(\Delta x)^2\sim t^\alpha$ with anomalous diffusion exponent $\alpha$.}
    \label{fig_visual}
\end{figure}

For constructing word embeddings, we use Word2vec's Skip-gram model with negative sampling (SGNS), which is one of the most widely used word embedding algorithms due its computational efficiency and ability to capture semantic relationships in a simple mathematical form \cite{mikolov2013distributed, mikolov2013efficient}. It uses a two-layer neural network to represent each word $i$ with two $D$-dimensional vectors called 'word vector' ($v_i$) and 'context vector' ($w_i$) such that a global cost function $C$, depending on all word vectors and all context vectors is (approximately) minimized. The objective of the Skip-gram model is to optimize the estimated empirical log-probabilities
\begin{equation}
    \log p (\,j\, |\,i\, ) ~\propto~ v_i \cdot w_j \label{eq_log-prob}
\end{equation}
of word $j$ occurring in the context of word $i$, using the  cost function $C$ defined as
\begin{equation}
    C = \frac{1}{L} \sum_{i=1}^{L} \sum_{j \in \mathcal{C} (i)} \log p (\,j\, |\,i\,), \label{eq_cost}
\end{equation}
where $L$ is the length of the text and $\mathcal{C} (i)$ is the linguistic context of word $i$. Following the recommendations of \cite{levy2015improving} and \cite{hamilton2016diachronic}, we set the dimension $D$ of the embedding space to $D=300$ and the context size to 2. Figure \ref{fig_visual}a visualizes a single embedding projected to 2 dimensions by t-sne \cite{van2008visualizing}, a non-linear dimension reduction method.

While constructing semantic trajectories of words from time-labelled co-occurrence data, one needs to account for two sources of arbitrariness in the process:  stochasticity and symmetry. The first, stochasticity, comes from the nature of modern data-efficient semantic embedding methods, such as Word2vec. Since these approaches use a neural network to find the best possible embedding, sampling techniques, initial weights, and the choice of numerical optimization method (such as stochastic gradient descent) lead to non-deterministic embeddings. Second, the cost $C$ itself exhibits multiple global minima associated with the same input data. In particular, the form of the estimated log-probabilities defined in (\ref{eq_log-prob}) implies that any transformation that leaves all dot products $ v_i \cdot w_j $ invariant results in the exact same cost $C$. The simplest example of such transformation is a linear rescaling of word vectors $v\mapsto \lambda v$ and inverse rescaling of context word vectors $w\mapsto \lambda^{-1} w$, but in principle, any invertible linear transformation of word vectors $v\mapsto R v$, together with (the transpose of) its inverse applied to context vectors $w\mapsto (R^{-1})^T w$ leaves the dot product invariant (see Methods for details).

However, an additional constraint comes from focusing solely on words (and not contexts) when constructing temporal trajectories: the constraint that ensures that the ambient embedding space does not shrink or expand over time, formalized as
\begin{equation}
    \Tr V = \sum_{i=1}^{D} \sigma_i^2 = \text{const}, \label{eq_fix_size}
\end{equation}
where $V$ is the $D\times D$ empirical covariance matrix of word positions (with $D=300$ being the embedding dimension) and $\sigma_i$ is the standard deviation of word positions along principal dimension $i$. As shown in Methods, this reduces the possible transformations $R$ to the orthogonal ones, obeying $R^{-1}= R^T$. Consequently, a single embedding is identified with an equivalence class containing $\{R v_i, R w_j\}$ with any orthogonal $R$.

We use this orthogonal freedom to define maximally representation-agnostic trajectories in two steps as shown in Figure \ref{fig_visual}b. First, we minimize the effect of stochasticity, described below, in any single time. Without stochasticity, any embedding starting from the same co-occurrence data would belong to the same equivalence class, and therefore it would be possible to find a transformation $R_j$ for each embedding $j$ such that they all numerically coincide. With stochasticity, however, different embedding realizations, starting from the same data, land in (slightly) different equivalence classes, and as a consequence, perfect alignment among them is not possible. The best one can do is to find a transformation for each embedding such that an overall distance measure between all embeddings is minimized (see Methods for details). This allows us to ”average out”, i.e., minimize the effect of stochasticity at any single timestep by taking the average of all aligned embeddings. Second, we align averaged embeddings corresponding to \emph{different} times in the same way: we find an orthogonal transformation that minimizes the distance between the embedding at subsequent times to construct word trajectories. Such word trajectories are thus maximally smoothened. Although this raises the question whether maximal smoothening washes away real phenomena, we will see that this smoothening procedure applied to random walk does not change measured observables such as the anomalous diffusion exponent $\alpha$ of the process. Figure \ref{fig_visual}c shows the measured semantic change of three selected words within a decade, projected to 2 dimensions, visualized over a static background.

\subsection{Semantic subdiffusion across languages}

Starting from word trajectories, we construct the ensemble of all trajectories and ask whether such an ensemble of trajectories obeys any robust statistical regularities. We focus on measuring the deviation of such trajectories from those of standard diffusing particles (i.e., a random walk), quantified by the \emph{anomalous diffusion exponent} $\alpha$, defined as \cite{bouchaud1990anomalous, metzler2000random, klages2008anomalous, metzler2014anomalous}

\begin{equation}
    | \Delta x |^2 \sim t^{\alpha}, \label{eq_alpha}
\end{equation}

where $\Delta x = x(t) - x(0)$ is the $D$-dimensional displacement vector of a word at time $t$ compared to its starting position at time $t=0$. In particular, we are interested in a potential deviation from various null-model trajectories that i) possess short memory, ii) are weakly interacting, iii) are driven by non-varying dynamical rules, and iv) step sizes or waiting times between steps are not overly heterogeneous. Such stochastic trajectories are characterized by an anomalous diffusion exponent $\alpha=1$. On the other hand, if either i), ii), iii), or iv) (or other conditions we do not discuss here) are violated, the process might belong to the anomalous diffusive regime with an observed exponent $\alpha\neq 1$ \cite{bouchaud1990anomalous, metzler2000random, klages2008anomalous, metzler2014anomalous}. Figure 1d illustrates subdiffusive trajectories, corresponding to various $\alpha<1$ exponents, generated by fractional Brownian motion (fBm) \cite{dieker2004simulation}. Note that we chose fBm for visualization purposes only; fBm generates trajectories with long-range temporal correlations, corresponding to the violation of point i) above . Actual semantic trajectories might be governed by (a mixture of) other underlying dynamical rules, as discussed below.

To measure the actual value of the anomalous diffusion exponent $\alpha$, we proceed from single trajectories to an ensemble of trajectories in two alternative ways: (i) we first average $|\Delta x|^2(t)$ over individual trajectories to obtain $\langle|\Delta x|^2(t)\rangle$ and then we fit the \emph{ensemble-average anomalous diffusion exponent}, $\langle \alpha \rangle$, based on $\langle|\Delta x|^2(t)\rangle\sim t^{\langle \alpha \rangle}$; (ii) alternatively, we first fit the anomalous diffusion exponent $\alpha$ to single trajectories and then we average over words to obtain the \emph{mean anomalous diffusion exponent} $\bar{\alpha}$. Although individual trajectories deviate considerably from the simple scaling behavior given by $|\Delta x|^2\sim t^{\alpha}$, somewhat surprisingly, their ensemble average, $\langle|\Delta x|^2(t)\rangle$, follows the scaling given by $t^{\langle \alpha \rangle}$ with high accuracy. This is illustrated in Figure \ref{fig_results}a, showing the squared displacement $|\Delta x|^2(t)$ of several individual English words as well as the ensemble average $\langle|\Delta x|^2(t)\rangle$ of all English words.

The grey cells of Table \ref{tab_exponents} list obtained exponents $\langle \alpha \rangle$ and $\bar{\alpha}$ for five languages, English, French, German, Italian, and Spanish. In all languages, both the ensemble-average anomalous diffusion exponent $\langle \alpha \rangle$ and the mean anomalous diffusion exponent $\bar{\alpha}$ are significantly lower than $\alpha = 1$, i.e., semantic trajectories follow subdiffusion. In particular, they follow subdiffision with an ensemble-average anomalous diffusion exponent $0.4 < \langle \alpha \rangle  < 0.5$ for all five languages. This is in strong contrast with a random walk generated using the same parameters (see Methods for details), corresponding to a fitted $\alpha\approx 1$, as shown by Figure \ref{fig_results}b.

\subsection{Comparison with randomized trajectories}

\begin{figure}
    \centering
    \includegraphics[width=\linewidth]{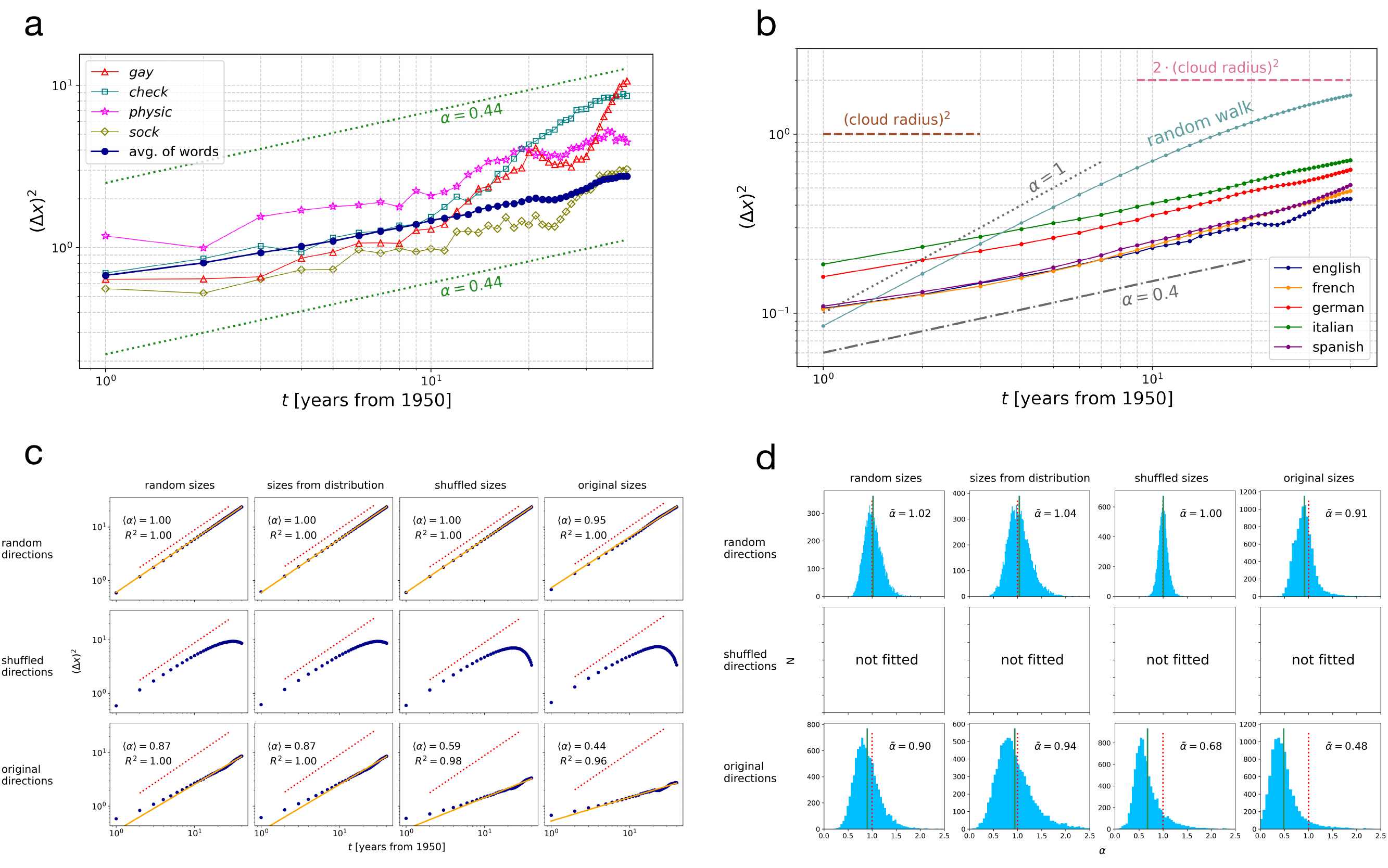}
    \caption{(a) Squared displacement $(\Delta x)^2$ of selected English words over time, along with average squared displacement over all words, shown in blue, which is well-approximated by $\langle (\Delta x)^2\rangle\sim t^{\langle \alpha \rangle}$ with ensemble-average anomalous diffusion exponent $\langle \alpha \rangle \approx 0.44$. (b) Average squared displacement of words over time in five languages, English, French, German, Italian, and Spanish. This is compared to an ensemble of random walkers under the same constraint given by a constant total word cloud size, defined by Eq. (\ref{eq_fix_size}). (c) Average squared displacement $\langle (\Delta x)^2\rangle$ of English words under various combinations of trajectory randomization methods. Columns and rows differ in the way step \emph{sizes} and \emph{directions} are randomized, respectively, see details in text. Keeping the original sequence of step directions is necessary to recover strongly subdiffusive behavior (bottom row), but not sufficient: the more information regarding step sizes are kept, the more its best fit diffusion exponent approaches its original value of $\langle \alpha \rangle\approx 0.44$. (d) Distribution of diffusion exponents $\alpha$ fitted to the trajectory of each word separately. Their average $\bar{\alpha}$ is close to the ensemble averages $\langle \alpha \rangle$ shown in c).}
    \label{fig_results}
\end{figure}

Given the robust observation that ensemble-average semantic trajectories of words follow subdiffusion with $\langle \alpha \rangle \approx 0.4-0.5$ across languages, one might ask the following two questions. (i) Is this an artifact of the diachronic alignment procedure? (ii) If not, what is behind the observed subdiffusion? In other words, what (combination of) microscopic models of stochastic collective dynamics might explain this macroscopic result? In the following, we focus on these two questions. We apply a series of randomization methods to the original trajectories, gradually removing temporal correlations in step sizes and step directions of individual trajectories to see which of these, if any, play a role behind subdiffusion. As shown in Figure \ref{fig_results}c and Table \ref{tab_exponents}, step sizes of a trajectory are randomized three different ways, which we call \emph{random sizes}, \emph{sizes from distribution}, and \emph{shuffled sizes}; step directions are randomized two ways, \emph{random directions}, \emph{shuffled directions} (see Methods), which, together with the original trajectories, gives three times four combinations. 

Figure \ref{fig_results}c and \ref{fig_results}d show the average squared displacement $\langle (\Delta x)^2\rangle$ of all English words, and the distribution of anomalous diffusion exponents $\alpha$ fitted to individual word trajectories separately, under all twelve combinations of trajectory randomization methods (including the original trajectories). The same plots for French, German, Italian, and Spanish are shown in the SI; fitted anomalous diffusion exponents $\langle \alpha \rangle$ and $\bar{\alpha}$ are listed for all five languages in Table \ref{tab_exponents}. The top left panel (random step sizes, random step directions) correspond to a random walk; the bottom right panel (original step step sizes, original step directions) corresponds to the original, non-randomized semantic trajectories. Comparing the results of original trajectories with the randomized trajectories (figure \ref{fig_results}c), conveys three important messages: (i) temporal alignment, applied to an ensemble of uncorrelated trajectories (top left panel), results $\langle\alpha\rangle=1$ (see top left panel), equivalent to that of uncorrelated trajectories \emph{without} alignment. (ii) Both correlations in step sizes and step directions are important factors behind subdiffusion. Neither of them alone can produce trajectories with an $\alpha$ lower than 0.8, yet the two combined gives $\alpha\approx 0.5$. (iii)
If step sizes do not vary significantly, the shuffling of the directions can alter the \emph{total} displacement along a trajectory only by a small amount, thus, the last data points in the middle row (shuffled directions) must be very close to the last data points of the corresponding panels of the last row (original directions).
On the other hand, shuffled directions do remove temporal correlations in step directions, making trajectories follow approximately diffusion-like behavior until the constraint on total displacement does not affect them. This can be clearly seen in the middle row of Figure \ref{fig_results}c; in Figure \ref{fig_results}d, we decided not to fit exponents to individual word trajectories in the middle row (shuffled directions) to avoid systematic bias in the exponents depending on the fitting range.

We further investigate a related effect, the effect of keeping the total size of the word \emph{cloud} constant, as formalized by Eq.(\ref{eq_fix_size}). Ensembles of randomized trajectories in Figure \ref{fig_results}c and \ref{fig_results}d do not obey this constraint; we therefore generated a random walk, corresponding to random step directions and random step sizes, with the additional constraint on keeping the total cloud size constant. This simulated trajectory is illustrated in figure \ref{fig_results}b ("random walk"). While the total displacement is limited by the size of the word could, this only appears to affect the random walk trajectory when the displacement reaches the radius of the cloud (and then converging to approximately to the square root of two times the radius, corresponding to two orthogonal vectors with lenght equals to the radius). This, along with the middle row of Figure \ref{fig_results}c, suggest that global constraints on total displacement do not cause the observed subdiffusive behavior.

\begin{table}
    \centering
    \begin{tabularx}{0.55\textwidth}{|c| *4{Y|} }
            \hline
            \multirow{2}{*}{\textbf{ENGLISH}} & \multicolumn{2}{c|}{random directions} & \multicolumn{2}{c|}{original directions} \\
            & \multicolumn{1}{c}{$\langle \alpha \rangle$} & \multicolumn{1}{c|}{$\bar{\alpha}$} & \multicolumn{1}{c}{$\langle \alpha \rangle$} & \multicolumn{1}{c|}{$\bar{\alpha}$} \\\hline
            random sizes & 1.00 & 1.02 & 0.87 & 0.90 \\\hline
            sampled sizes & 1.00 & 1.04 & 0.87 & 0.94 \\\hline
            shuffled sizes & 1.00 & 1.00 & 0.59 & 0.68 \\\hline
            original sizes & 0.95 & 0.91 & \cellcolor{gray!30}\textbf{0.44} & \cellcolor{gray!30}\textbf{0.48} \\\hline
        \end{tabularx}
        \vspace{0.4cm}
        
        \begin{tabularx}{0.55\textwidth}{|c| *4{Y|} }
            \hline
            \multirow{2}{*}{\textbf{FRENCH}} & \multicolumn{2}{c|}{random directions} & \multicolumn{2}{c|}{original directions} \\
            & \multicolumn{1}{c}{$\langle \alpha \rangle$} & \multicolumn{1}{c|}{$\bar{\alpha}$} & \multicolumn{1}{c}{$\langle \alpha \rangle$} & \multicolumn{1}{c|}{$\bar{\alpha}$} \\\hline
            random sizes & 1.00 & 1.03 & 0.90 & 0.93 \\\hline
            sampled sizes & 1.00 & 1.05 & 0.90 & 0.98 \\\hline
            shuffled sizes & 1.00 & 1.00 & 0.57 & 0.68 \\\hline
            original sizes & 0.98 & 1.02 & \cellcolor{gray!30}\textbf{0.49} & \cellcolor{gray!30}\textbf{0.64} \\\hline
        \end{tabularx}
        \vspace{0.4cm}
        
        \begin{tabularx}{0.55\textwidth}{|c| *4{Y|} }
            \hline
            \multirow{2}{*}{\textbf{GERMAN}} & \multicolumn{2}{c|}{random directions} & \multicolumn{2}{c|}{original directions} \\
            & \multicolumn{1}{c}{$\langle \alpha \rangle$} & \multicolumn{1}{c|}{$\bar{\alpha}$} & \multicolumn{1}{c}{$\langle \alpha \rangle$} & \multicolumn{1}{c|}{$\bar{\alpha}$} \\\hline
            random sizes & 1.00 & 1.02 & 0.81 & 0.84 \\\hline
            sampled sizes & 1.00 & 1.03 & 0.81 & 0.87 \\\hline
            shuffled sizes & 1.00 & 1.00 & 0.51 & 0.57 \\\hline
            original sizes & 0.97 & 0.98 & \cellcolor{gray!30}\textbf{0.41} & \cellcolor{gray!30}\textbf{0.48} \\\hline
        \end{tabularx}
        \vspace{0.4cm}
        
        \begin{tabularx}{0.55\textwidth}{|c| *4{Y|} }
            \hline
            \multirow{2}{*}{\textbf{ITALIAN}} & \multicolumn{2}{c|}{random directions} & \multicolumn{2}{c|}{original directions} \\
            & \multicolumn{1}{c}{$\langle \alpha \rangle$} & \multicolumn{1}{c|}{$\bar{\alpha}$} & \multicolumn{1}{c}{$\langle \alpha \rangle$} & \multicolumn{1}{c|}{$\bar{\alpha}$} \\\hline
            random sizes & 1.00 & 1.02 & 0.77 & 0.79 \\\hline
            sampled sizes & 1.00 & 1.02 & 0.78 & 0.81 \\\hline
            shuffled sizes & 1.00 & 1.00 & 0.51 & 0.55 \\\hline
            original sizes & 0.94 & 0.96 & \cellcolor{gray!30}\textbf{0.40} & \cellcolor{gray!30}\textbf{0.44} \\\hline
        \end{tabularx}
        \vspace{0.4cm}
        
        \begin{tabularx}{0.55\textwidth}{|c| *4{Y|} }
            \hline
            \multirow{2}{*}{\textbf{SPANISH}} & \multicolumn{2}{c|}{random directions} & \multicolumn{2}{c|}{original directions} \\
            & \multicolumn{1}{c}{$\langle \alpha \rangle$} & \multicolumn{1}{c|}{$\bar{\alpha}$} & \multicolumn{1}{c}{$\langle \alpha \rangle$} & \multicolumn{1}{c|}{$\bar{\alpha}$} \\\hline
            random sizes & 1.00 & 1.03 & 0.93 & 0.97 \\\hline
            sampled sizes & 1.00 & 1.06 & 0.93 & 1.02 \\\hline
            shuffled sizes & 1.00 & 1.01 & 0.63 & 0.78 \\\hline
            original sizes & 0.93 & 0.94 & \cellcolor{gray!30}\textbf{0.49} & \cellcolor{gray!30}\textbf{0.66} \\\hline
        \end{tabularx}
        \vspace{0.4cm}
        
        \caption{Ensemble-average anomalous diffusion exponent $\langle \alpha \rangle$ and the average $\bar{\alpha}$ of the anomalous diffusion exponents of all words, under various combinations of trajectory randomization methods, for all five languages. Grey cells show the exponents of the original, non-randomized trajectories. Note that the exponents for English are extracted from Figures \ref{fig_results}c and \ref{fig_results}d; for the other four languages, the analogous Figures are included in the SI.}
        \label{tab_exponents}
\end{table}

\section{Discussion}

Cumulative culture is arguably one of the most distinct characteristic feature of human behavior. Understanding "laws" that govern cultural evolution is a crucial step towards understanding Homo Sapiens itself. Here we study cultural evolution through the evolution of \emph{meaning} of words, as formalized by the distributional hypothesis: "a word is characterized by the company it keeps" \cite{lenci2008distributional}. We rely on large time-labelled corpora, Google Ngram, available in several languages. Cultural evolution is notoriously difficult to study without making a large number of subjective assumptions and interpretations. 
One of the main contribution of this work is that it tries to make underlying assumptions (that are in the form of mathematical formalizations and their interpretation) as explicit as possible.

Our data processing and analysis pipeline consist of three phases. First, semantic relations between all words are extracted through a state-of-the-art implementation of the distributional hypothesis: Word2vec, trained with the so-called Skip-gram with negative sampling method. This algorithm provides a high-dimensional embedding of all words such that pairwise distances reflect semantic similarity. Apart from being fast and data efficient, it is also well anchored in human language representation through psycholiguistic studies.  Second, to extract evolutionary trajectories, embeddings at different times need to be weaved together. This is a highly non-trivial process since the mapping between embedding and co-occurrence statistics is degenerate: many embeddings are consistent with the same co-occurrence data. When constructing trajectories, we need to break this symmetry: one specific embedding from each equivalence class needs to be chosen. We jointly construct equivalence classes and choose one specific representative of each class, both informed by the dynamics. In particular, we choose trajectories to be maximally smoothened. As we show, this maximal smoothening, applied to an ensemble of diffusing particles, does not alter the anomalous diffusion exponent $\alpha$ of the process, suggesting that it would not alter $\alpha$ significantly when actual semantic trajectories are considered, either. The third phase of the process is the comparison of the ensemble of semantic trajectories of words with various randomized counterpart ensembles.
The randomization method we consider dissects various temporal correlations in semantic trajectories by randomizing step directions and step sizes separately, while still applying the same method for temporal alignment (i.e. trajectory smoothening).
With all this, we seek to answer the following questions: (i) is there any robust statistical regularity regarding the ensemble of actual semantic trajectories of words? (ii) If yes, what might be the reason? What microscopic dynamical rules can explain the observed macroscopic (ensemble-level) statistical regularities? This work provides an answer to the first question: semantic trajcetories are very different from ordinary random walk (diffusion); semantic trajectories are strongly subdiffusive. In particular, actual semantic trajectories follow subdiffusion with an anomalous diffusion exponent $\alpha\approx 0.4$ to $0.5$, in strong contrast with a random walk belonging to the $\alpha=1$ class.

We point out here that short-range temporal correlations in trajectories, not extremely inhomogeneous step sizes, or weak correlations between trajectories do not cause the resulting trajectories to deviate from $\alpha=1$. Instead, subdiffusion can be explained by qualitatively different microscopic dynamical rules. These include (i) long temporal correlations in trajectories, (ii) extremely inhomogeneous step size distributions, (iii) stochastic dynamics of "jamming" (overly densely packed) particles, (iv) changing average step sizes over time (corresponding to a changing diffusion coefficent), (v) diffusion in disordered media, and many others. Although investingating possible combinations of these microsopic dynamics as explanations of semantic subdiffusion is a subject of future work, we can at least exclude some of them based on our results. Step size distribution ("Brownian vs Levy flight") and even correlations in step sizes do not seem to contribute to subdiffusion at all. Correlations in step directions explain some but far from all: ensembles of trajactories where step sizes are randomized but step directions are kept still exhibit $\alpha\approx 0.8-0.9$ in contrast with actual trajectories that follow $\alpha\approx 0.4-0.5$ (that include correlations both among step sizes and directions, and possibly cross-correlations between these categories too).

Subdiffusive behavior has been observed in various within-cell processes, such as in the stochastic trajectory of messenger RNA inside living E. coli cells, $\alpha\approx 0.7$, \cite{golding2006physical}, channel proteins in the membranes of living cells, $\alpha\approx 0.9$, \cite{weigel2011ergodic}, and telomeres within eucaryotic cell nuclei, $\alpha\approx 0.3$, \cite{bronstein2009transient}. As in the biological examples above, subdiffusive behavior of semantic change raises questions both at a mechanistic, proximal level (what microscopic dynamical rules underlie subdiffusion?) and at an evolutionary, distal level (is subdiffusion adaptive or is it a consequence of physical-informational constraints? If it is adaptive, what is it \emph{for} and what selection pressures led to its emergence?). Although answering these questions is outside the scope of this current paper, we hope that our results stimulate extensive future discussions with a strong interdisciplinary focus along the lines of the questions listed above.

\section{Methods}

\begin{figure}
    \centering
    \includegraphics[width=0.75\linewidth]{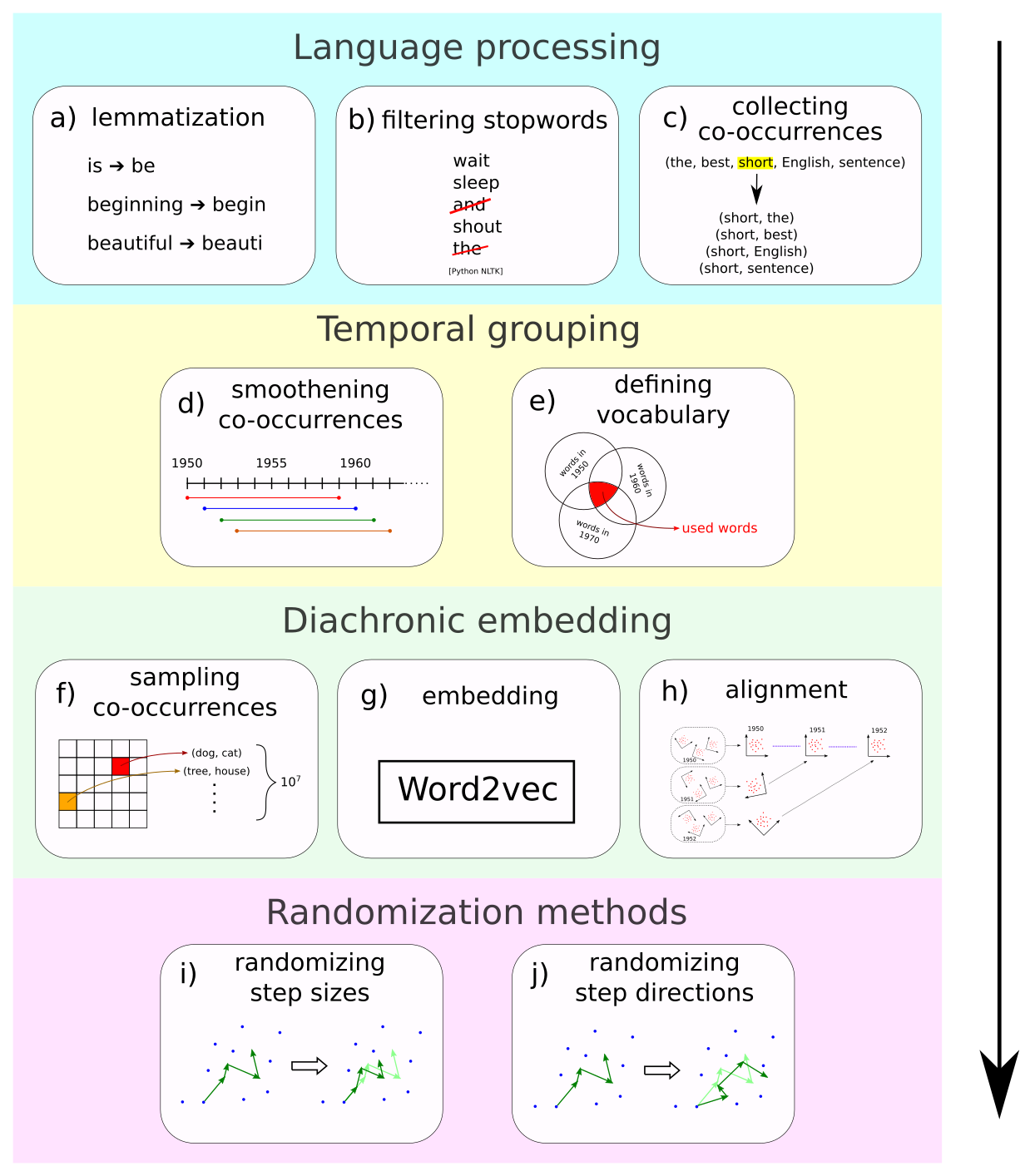}
    \caption{Key steps of the pipeline that creates diachronic embeddings from raw google ngram data.
    (a) Lemmatization of words.
    (b) Filtering stopwords.
    (c) Counting co-occurrences with context window of lenght 2.
    (d) Smoothening co-occurrences by applying a sliding time window to co-occurrence counts.
    (e) Vocabulary, i.e., the final set of lemmas we include in our analysis, is defined by the lemmas that occur at least 100 times at all time windows. 
    (f) Training dataset was created by subsampling all co-occurrences in order to have a constant number of co-occurrences $N_c=10^7$ at each time window.
    (g) Word embedding by Word2vec with Skip-gram.
    (h) Alignment of different embeddings (both within and across years).
    (i) Randomization of the temporal order of step \emph{sizes} of each word trajectory separately.
    (j) Randomization of the temporal order of step \emph{directions} of each word trajectory separately.}
    \label{fig_pipeline}
\end{figure}

To construct semantic trajectories of words, we use the downloadable version of the Google Books Ngram Viewer database, version 2 \footnote{available to download at https://storage.googleapis.com/books/ngrams/books/datasetsv2.html} as time-labelled corpora available in multiple languages \cite{michel2011quantitative,lin-etal-2012-syntactic}. In this work, we included five languages from the Indo-European family with the most available data, English, French, German, Spanish, and Italian. Ngram data is processed and analysed in four steps, summarized in Figure \ref{fig_pipeline}. 

\subsection{Language processing}

We used the snowball stemmer of the natural language toolkit (nltk) module of Pyton \cite{bird2009natural} to lemmatize words, and then we filtered out stopwords \cite{schutze2008introduction}, before collecting all co-occurrences within a window size 2, as suggested in \cite{levy2015improving} and \cite{hamilton2016diachronic}, illustrated by Figure \ref{fig_pipeline}a,b,c.

\subsection{Temporal grouping} \label{subsubsec_temporal}

Since temporal annotation of ngram occurrences follow a yearly resolution, we first collected co-occurrences by year, and then we applied a 10-year sliding window to increase vocabulary size and to smoothen data. Although setting the sliding window size to 10 years is an arbitrary choice, we show in the SI that as window size increases gradually from 1 year to 10 years, the exponent $\alpha$ approximately converges to its value at 10 years. We then filtered out words that do not appear at least 100 times in each time window. This process is shown in Figure \ref{fig_pipeline}d,e.

\subsection{Measuring semantic distance} \label{subsec_distance}

In this paper, we self-consistently define semantic distance $|x-y|$ between vectors $x$ and $y$ in the D-dimensional embedding space as Euclidean distance, $|x-y|^2=\sum_{i=1}^D (x_i-y_i)^2$. This is a departure from the most commonly used semantic similarity measure, cosine similarity $|x-y|_{\text{cos}}$, which is favored because it accounts for the rescaling symmetry, $x\to \lambda x$, by setting $|\lambda_1 x-\lambda_2 y|_{\text{cos}}=|x-y|_{\text{cos}}$ \cite{landauer1998introduction, mikolov2013efficient}. We argue, however, that finding the most general class of underlying symmetries and measuring distance separately is a more principled procedure, especially when semantic distances have numerical and not just ordinal meaning. In particular, we choose Euclidean distance for the following reasons.

1. In the Skip-gram model, the difference of two word vectors, $v_i - v_j$, has a semantic meaning. By interpreting the space of context words as a dual space (see SI for details), the meaning of $v_i - v_j$ can be understood by projecting it to various context words, akin to the definition of a linear functionals in dual space, 

\begin{equation}
\begin{split}
    (v_i - v_j) \cdot w_k &= v_i \cdot w_k - v_j \cdot w_k\\
    & \propto \log p ( \,k\, | \,i\, ) - \log p ( \,k\, | \,j\, ) \\
    & \propto \log \left( \frac{p ( \,k\, | \,i\, )}{p ( \,k\, | \,j\, )} \right).
\end{split}
\end{equation}

Since this result is valid for any context word $k$, $v_i - v_j$ measures the relative log-probability of word $i$ versus word $j$ occurring in the context of word $k$, for all contexts $k$. This is in line with the starting hypothesis of distributional semantics, which states that two words have a similar meaning if their co-occurrence statistics is similar. 

2. Euclidean distance is a \emph{metric}, obeying triangle inequality, commutativity, and other axioms, whereas other measures, such as cosine similarity is not. This is important when semantic trajectories are interpreted as trajectories in metric spaces, for example, by measuring their squared displacement over time, $|\Delta x|^2(t)=|x(t)-x(t=0)|^2$.

3. As discussed above, embedding dimensions are arbitrary. They need to be aligned for meaningful comparison across multiple embeddings (regardless of it coming from the same co-occurrence data or a different one). Here we follow the most commonly used method, called the Procrustes algorithm \cite{schonemann1966generalized}, that finds an orthogonal transformation that minimizes the total Euclidean distance between corresponding points in the two embeddings. Measuring semantic distance as Euclidean distance is consistent with this alignment method.

4. As a consequence of keeping the size of the word cloud constant, defined by Eq. (\ref{eq_fix_size}), two embeddings belong to the same equivalence class if they can be orthogonally transformed to each other, discussed above and shown below in Methods. Orthogonal transformations are defined by keeping Euclidean distance, also consistent with our definition.

\subsection{Diachronic embedding}

We first subsample the co-occurrences corresponding to each time window to eliminate systematic sample size bias (the amount of data included in the Google Ngram database steadily increases with time in all five languages). We set this sample size to $N_c=10^7$ co-occurrences, set by the first (few) time windows with the least amount of data. We observe that $N_c=10^7$ sets a good tradeoff between sample size (and thus trajectory noise), vocabulary size, and trajectory lenght for this database. 

After generating the $D=300$ dimensional embeddings for each time window $M=80$ times by Word2vec with Skip-gram, we first align all $M$ embeddings within a time window. Then, word positions across embeddings are averaged to obtain a more robust estimate of semantic positions for each word at a given time window. These average embeddings, one for each time window, are then aligned across time, as shown by Figure \ref{fig_pipeline}h. 

The alignment of two embeddings consists of two steps. First, one has to define the class of transformations that keep semantic relationships invariant, and second, the transformation within this class that minimizes the difference between the two embeddings has to be selected. For the first step, we find (see below for a proof) that the requirement of keeping the total size of the word cloud constant, as defined by Eq. (\ref{eq_fix_size}), restricts all linear transformations to those that are orthogonal. This is also in line with both measuring semantic distance as euclidean distance, as discussed above, and also with most commonly used aligment method called the orthogonal Procrustes method, which finds that the best transformation $R$ that accounts for this rotational (orthogonal) freedom is given by
\begin{equation}
    R = \arg \min_{Q \in \mathrm{O} \left(D\right)} || W' Q - W ||_F \label{eq_procrustes}
\end{equation}
where the rows of the $N \times D$ matrix $W'$ and $W$ contain the word vectors of the aligned and reference embedding, respectively, and $|| \cdot ||_F$ is the Frobenius matrix norm. We use the analytical solution \cite{schonemann1966generalized} for $R$ to find optimal alignments both within a time window and among time windows.

\subsection{Constant word cloud size defines an orthogonal embedding symmetry}

Word cloud size, defined by Eq. (\ref{eq_fix_size}), is written as
\begin{equation}
     \Tr V \left( W \right) =  \sum_{j = 1}^{D} \mathop{\mathrm{Var}}_i \left( W_{ij} \right) \label{eq_size}
\end{equation}

where $V(W)$ is the empirical covariance matrix of word positions, extracted from $W$, the $N \times D$ matrix with its rows corresponding to the $D$-dimensional word vectors of all $N$ words, and $\mathop{\mathrm{Var}}_i \left( W_{ij} \right)$ is the variance of the word positions along dimension $j$. Using the definition of variance,

\begin{equation}
\begin{split}
    \Tr V \left( W \right) &= \sum_{j = 1}^{D} \mathop{\mathrm{Var}}_i \left( W_{ij} \right) = \sum_{j=1}^D \left[ \frac{1}{N} \sum_{i=1}^{N} W_{ij} W_{ij} - \left( \frac{1}{N} \sum_{i=1}^{N} W_{ij} \right)^2 \right]  \\
    & = \frac{1}{N^2} \sum_{j = 1}^D \sum_{i=1}^{N} \left[ N W_{ij} W_{ij} - W_{ij} \sum_{k=1}^{N} W_{kj} \right].
\end{split}
\end{equation}

The size of the transformed word cloud $W'$, defined as $W'=RW$ is

\begin{equation}
\begin{split}
    \Tr V \left( W' \right) &= \frac{1}{N^2} \sum_{j = 1}^D \sum_{i=1}^{N} \left[ N \sum_{l=1}^D W_{il} R_{lj} \sum_{m=1}^D W_{im} R_{mj} - \sum_{l=1}^D W_{il} R_{lj} \sum_{k=1}^{N} \sum_{m=1}^D W_{km} R_{mj} \right] \\
    &= \frac{1}{N^2} \sum_{j, l, m = 1}^D \sum_{i=1}^{N} R_{lj} R_{mj} \left[ N W_{il} W_{im} - W_{il} \sum_{k=1}^{N} W_{km} \right].
\end{split}
\end{equation}

The difference between $\Tr V \left( W \right)$ and $\Tr V\left( W' \right)$ is

\begin{equation}
    \Tr V \left( W \right)- \Tr V \left( W' \right) = \frac{1}{N^2} \sum_{l, m = 1}^D \sum_{i=1}^{N} \left( \delta_{ml} - \sum_{j=1}^D R_{lj} R_{mj} \right) \left[ N W_{il} W_{im} - W_{il} \sum_{k=1}^{N} W_{km} \right].
\end{equation}

The transformation $R$ satisfies the requirement of constant cloud size $\Tr V \left( W \right) - \Tr V \left( W' \right) = 0$ only if 

\begin{equation}
\sum_{j=1}^D R_{lj} R_{mj} = \delta_{ml},
\end{equation}

also written in matrix form as $R R^T = \boldsymbol{1}$, i.e., $R$ needs to be orthogonal.

\subsection{Trajectory randomization methods}

Figure \ref{fig_pipeline}i and j illustrates the two types of trajectory randomization methods we use in this paper to generate the results shown by Figure \ref{fig_results}c,d and Table \ref{tab_exponents}: randomization of step sizes and randomization of step directions. Within each type, "randomization" refers to drawing step sizes and directions from various distributions constructed from the original trajectories, described as follows.

\emph{Random sizes}: step sizes were sampled from a normal distribution with mean and standard deviation corresponding to that of the step sizes of the original trajectory; \emph{sizes from distribution}: step sizes were sampled from the set of all step sizes corresponding to all words; \emph{shuffled sizes}:  step sizes were sampled from the set of step sizes corresponding to the same word; in other words, the temporal order of step sizes of a trajectory has been shuffled; \emph{original sizes}: step size for every word at every time step was set to its original value.

\emph{Random directions}: directions were sampled uniformly over a D-dimensional sphere; \emph{shuffled directions}: the temporal order of step directions of a trajectory has been shuffled; \emph{original directions}: the direction for every word at every time step was set to its original value.

\section*{Acknowledgements}

The authors thank Douglas Moore, Yanbo Zhang, Jake Hanson, András Szántó, József Venczeli, and Péter Pollner for useful discussions at various stages of the project. 
This project has received funding from the European Union’s Horizon 2020 research and innovation programme under grant agreement no. 101021607
and was partially supported by the National Research, Development and Innovation Office under grant no. K128780
and  by the the European Union project RRF-2.3.1-21-2022-00004 within the framework of the Artificial Intelligence National Laboratory
and by the ÚNKP-21-3 New National Excellence Program of the Ministry for Innovation and Technology from the source of the National Research, Development and Innovation Fund.

\section*{Data availability}

Python code is available at \verb|https://github.com/abogdan271/histwords|.

\section*{Author contributions}

All authors designed the project, worked out the mathematical formalisms, and wrote the paper. Data analysis and simulations by BA.

\bibliographystyle{unsrt}

\end{document}